\documentstyle[psfig]{ioplppt}
\begin{document}
\eqnobysec   %JPB
\jl{2}    %JPB
%\twocolumn[
%\draft    %JPB

\title{Counterintuitive transitions in multistate curve crossing
involving linear potentials}[Counterintuitive transitions in curve
crossing]

\author{ V A Yurovsky, A Ben-Reuven, P S Julienne\dag, Y B Band\ddag}

\address{School of Chemistry, Tel Aviv University, Tel Aviv 69978,
Israel}

\address{\dag\ Atomic Physics Division, PHYS A-167, National Institute
of Standards and Technology, Gaithersburg, MD 20889, USA}

\address{\ddag\ Departments of Chemistry and Physics, Ben Gurion
University, Beer Sheva 84105, Israel}

\date{\today}
%\widetext

\begin{abstract} Two problems incorporating a set of horizontal
linear potentials crossed by a sloped linear potential are
analytically solved and compared with numerical results: (a) the case
where boundary conditions are specified at the ends of a finite
interval, and (b) the case where the sloped linear potential is
replaced by a piecewise-linear sloped potential and the boundary
conditions are specified at infinity.  In the approximation of small
gaps between the horizontal potentials, an approach similar to the one
used for the degenerate problem (Yurovsky V A and Ben-Reuven A 1998
\jpb {\bf 31} 1) is applicable for both problems.  The resulting
scattering matrix has a form different from the semiclassical result
obtained by taking the product of Landau-Zener amplitudes.
Counterintuitive transitions involving a pair of successive crossings,
in which the second crossing precedes the first one along the
direction of motion, are allowed in both models considered here.

\end{abstract}
%\maketitle    %JPB
%]
%\narrowtext
\section{Introduction}

Transitions in multistate curve crossing may be represented
intuitively as a sequence of two-state crossings and avoided
crossings.  In the absence of turning points near the crossings, one
would expect that the crossings should occur in the causal ordering of
the crossing points along the direction of motion (see dashed arrow in
figure 1).  It is, however, known from quantum close-coupling
calculations that certain {\it counterintuitive} transitions may also
be allowed \cite{YB98,NWJ97,S96}, in which the causal arrangement may
be broken, letting the second crossing point precede the first one
with respect to the direction of motion (see solid arrow in figure 1).
Such transitions are generally forbidden in analytical semiclassical
theories of multistate curve crossing.

The concept of counterintuitive transitions \cite{YB98,NWJ97,S96}
has recently received some attention in the theory of cold atom
collisions, in particular regarding the problem of incomplete optical
shielding (or suppression) of loss-inducing collisions (see \cite
{NWJ97,S96,YB97} and references therein).  Optical shielding of a
colliding pair of cold atoms is attained by subjecting an atom to a
laser field with the laser frequency shifted to the blue of an
asymptotic atomic resonance frequency.  The laser field couples the
ground molecular state to a repulsive excited molecular state (which
correlates asymptotically to the state in which one of the atoms is
excited).  According to the ordinary (single crossing) Landau-Zener
(LZ) theory \cite{LZ}, the radiative coupling forms a repulsive
barrier which diverts and reflects the atoms approaching each other in
their ground states.  This theory predicts an exponential decrease of
the penetration (transmission) probability as the laser power is
increased.  Experiments indicate, however, that this shielding
efficiency saturates at a certain ``hangup'' value, in which the
transmission probability stays finite.

In some situations the optical shielding effect can be explained
by a semiclassical multiple-crossing model, associated with a pair of
transitions of the intuitive kind involving partial-wave channels
\cite{YB97}.  But in other situations, in which the former transitions
are impeded by various constraints (e.g., centrifugal barriers), it is
possible, as demonstrated by close-coupling calculations \cite{NWJ97},
to attribute the incomplete shielding effect to transitions of the
counterintuitive type.

In semiclassical approaches, transition amplitudes in multistate
systems are usually constructed from products of single-crossing (or
non-crossing) LZ amplitudes \cite{ChildSM,Nakamura}.  As already
stated, counterintuitive transitions are forbidden in such approaches.
Even in exactly-soluble models, such as the Demkov-Osherov model
\cite{DO67}, in which the semiclassical theory provides exact
transition amplitudes, counterintuitive transitions are forbidden.
This conclusion holds for non-degenerate channel potentials.  In a
recent publication \cite{YB98}, the Demkov-Osherov model has been
extended to the case in which some of the horizontal channel
potentials are degenerate.  A major observation of that work is that
an abrupt change occurs in the transition amplitudes, as the gap
between two such parallel potentials narrows to zero, making all
transitions possible.  Also, in this limit, the transition amplitudes
are no longer representable as a products of single-crossing LZ
amplitudes.  The range over which the transition occurs seems to
diverge on approaching degeneracy.

The Demkov-Osherov model is rather unusual, requiring a set of
flat (horizontal) parallel potentials crossing a single linear sloped
potential.  All potentials are assumed to retain these properties to
infinity, disregarding standard boundary conditions used in scattering
theory.  This peculiar property, combined with the observations on the
passage to degeneracy, have led us to inquire whether a modification
of the potentials at the far wings, away from the crossing region, may
lead to a correction of the Demkov-Osherov results, making the
counterintuitive transitions allowable when the potential gap between
the flat channels becomes sufficiently small.

We have found that this is indeed the case in the two modified
models we have solved.  In one model (described in section 2 below),
the domain of the model is truncated, confining it to a finite range,
with boundary conditions defined at its edges.  In the other model
(described in section 3), the single-sloped potential is replaced by a
piecewise-linear potential (see figure 1), constructed of three
connected segments (one finite central segment at the transition
range, and two semi-infinite segments in the wings).  We show here,
with the help of an analytical perturbation theory, that both models
allow for counterintuitive transitions.  The probability of these
transitions diminishes as the gap between the adjacent horizontal
potentials is increased.  These results, derived from the analytical
theory, are compared with numerical solutions of the associated
quantum close-coupling equations in section 4.

\section{Truncated linear problem} \label{TRUNLIN}

\subsection{Statement of the problem} \label{TRUNLINPS}

\begin{figure}

\psfig{figure=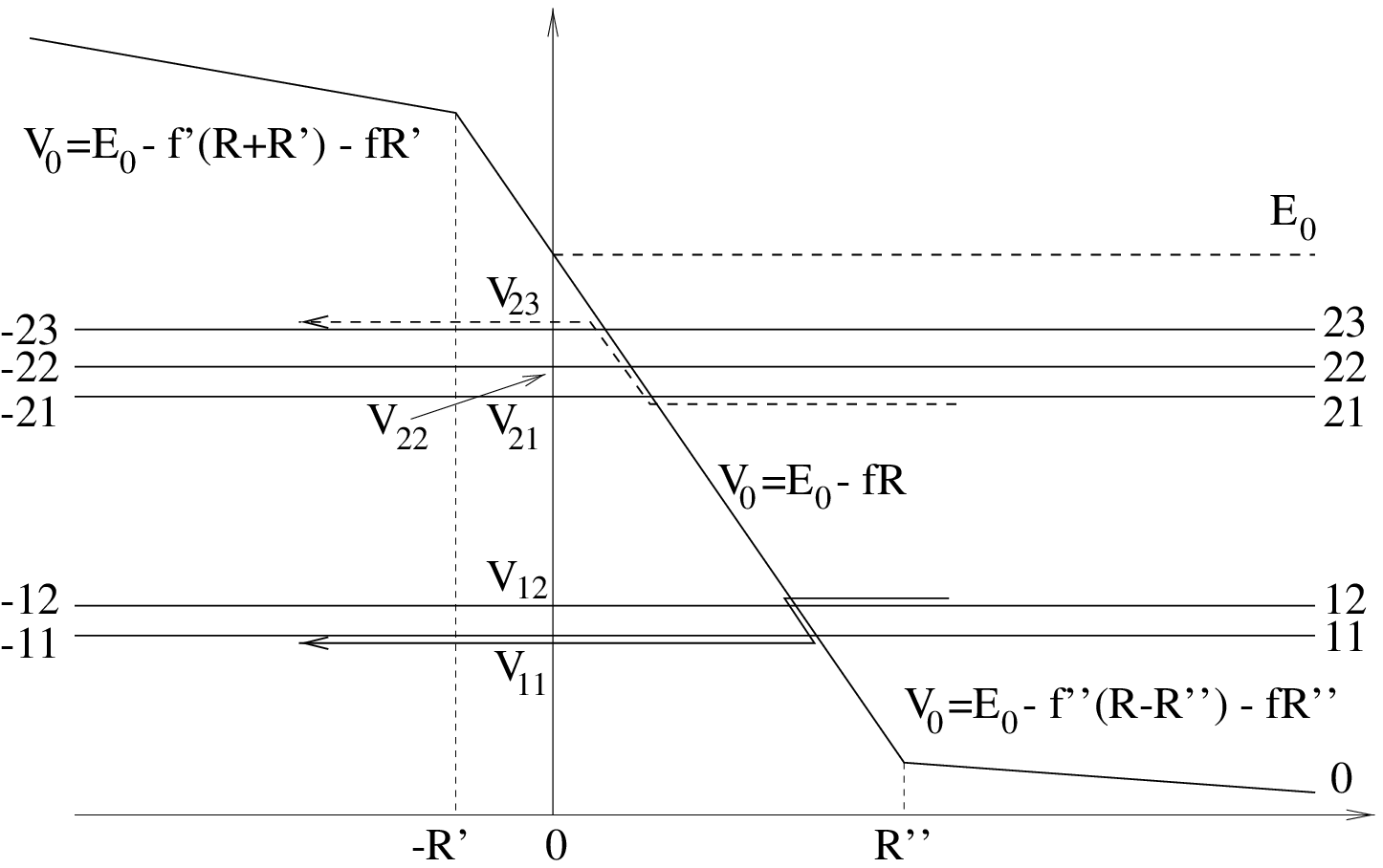}

\caption{Schematic illustration of the model of a sloped
piecewise-linear potential crossing a set of ($n=m=2$ here)
quasi-degenerate groups of horizontal potentials ($d_{1}=2$, $d_{2}=3$
here). Negative numbers denote transmission channels for waves
entering from the right. The truncated linear model involves
only the finite interval between $-R^\prime $ and $R^{\prime\prime}$.
 Dashed and solid
arrows show intuitive and counterintuitive transitions,
respectively.  \label{FigPot}}

\bigskip

\psfig{figure=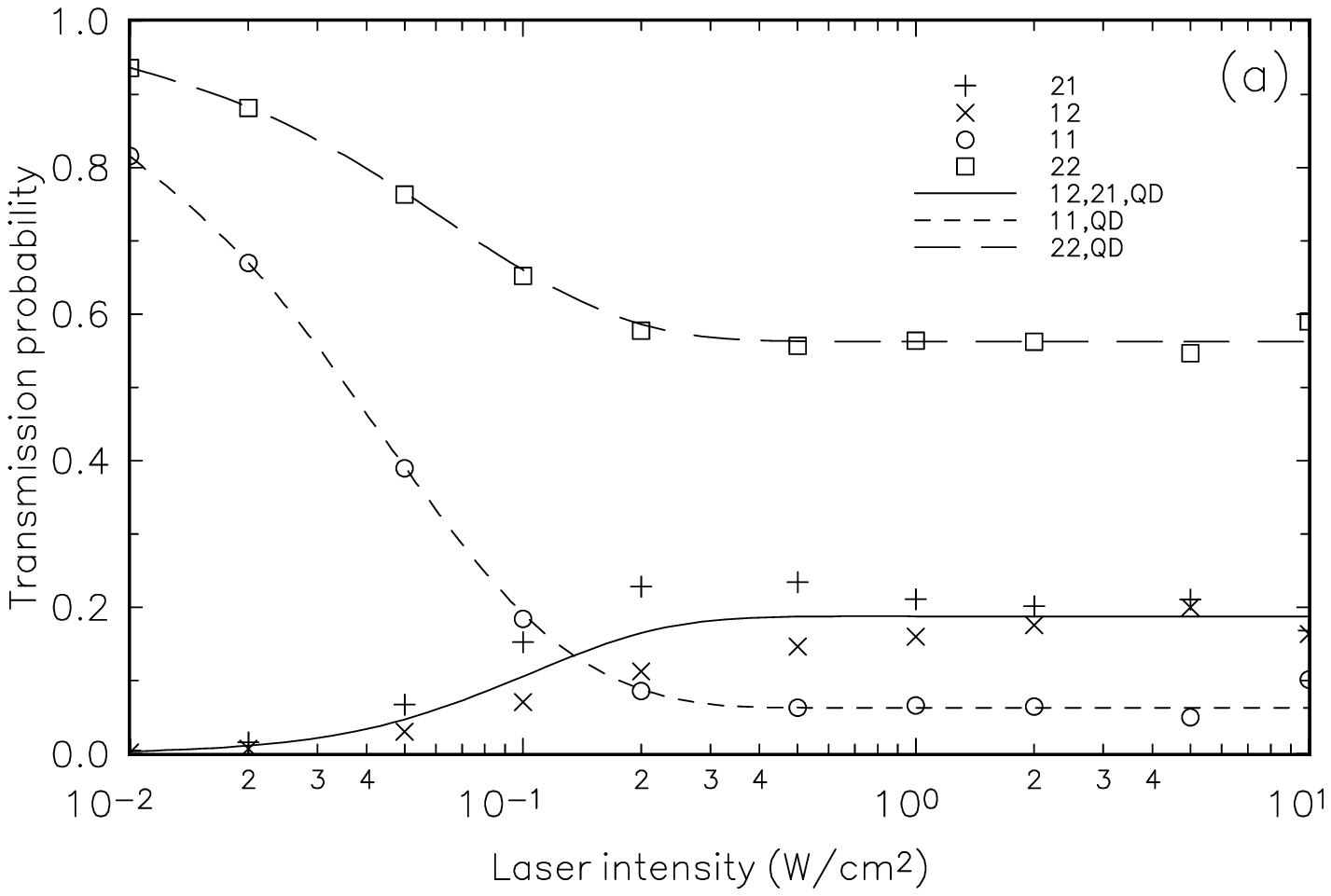}
\end{figure}

Consider a sloped linear potential crossing a set of
horizontal potentials (figure  \ref{FigPot}), bunched into some
quasi-degenerate groups. (The criteria defining quasi-degeneracy
will be specified below.) The case of exact degeneracy may be
reduced to the non-degenerate problem in the manner described in
\cite{YB98} and is therefore not considered here.

Let us denote as $|0\rangle $ the (internal) channel state with the
sloped potential $V_{0}$, and as  $|j\nu \rangle $ the channels with
 horizontal
potentials $V_{j\nu }$, where $1\le j\le m$ denotes a group of quasi
-degenerate
states and $1\le \nu \le d_{j}$  denotes a state within the group.
 The states are
arranged so that $V_{j\nu }<V_{j\nu ^\prime }$  for $\nu <\nu ^\prime
 $ and $V_{j\nu }<V_{j^\prime \nu ^\prime }$  for $j<j^\prime $  and
 all
$\nu $ and $\nu ^\prime $. The origin on the external coordinate axis
 $R$ is chosen as
the classical turning point on the sloped potential, so that $V_{0}
=E_{0}-
fR$, where $E_{0}$  is the total collision energy and $f$ is the
 repulsive
force. The collision energy also determines the highest open channel,
with $V_{nd_{n}}<E_{0}\left( n\le m\right) $. The problem is
 considered with boundary
conditions defined on a finite interval $-R^\prime <R
<R^{\prime\prime}$.

Substitution of the total wavefunction $\Psi $ in the form
\begin{equation}
\Psi =\sum\limits^{m}_{j=1}\sum\limits^{d{ } _{j}}_{\nu =1} a_{j\nu
 }\left( R\right)  |j\nu \rangle  + b\left( R\right)  |0\rangle
\end{equation}
into the Schr\"odinger equation leads to the set of close-coupling
equations for the coefficients $a_{j\nu }\left( R\right) $ and
 $b\left( R\right) $,\begin{equation}
\eqalign{
- {1\over 2\mu } {\partial ^{2}a{ } _{j\nu }\over \partial R{ } ^{2}}
+V_{j\nu }a_{j\nu }+g_{j\nu }b=E_{0}a_{j\nu }\qquad \left( 1\le j\le
 m\right)  \nonumber\\
- {1\over 2\mu } {\partial ^{2}b\over \partial R{ } ^{2}} - f R b
+\sum\limits^{m}_{j=1}\sum\limits^{d{ } _{j}}_{\nu =1} g_{j\nu
 }a_{j\nu }=0 .
}
\end{equation}(Atomic units are used here and in what follows.)
Here $\mu $ is the reduced mass, and the coupling constants $g_{j\nu
 }$  are
assumed real and $R$-independent. Without loss of generality, we can
also assume that the horizontal potential channels are not coupled
directly to each other (see  \cite{DO67}).

The solution presented in \cite{YB98} for the non-degenerate
case is applicable to the system discussed here if all the following
conditions hold:
\begin{equation}
\fl R^\prime ,R^{\prime\prime}\gg \max \left(  {g{ } ^{2}_{j\nu
 }\over f\left( E_{0}-V_{j\nu }\right) },\quad {E_{0}-V{ } _{j\nu
 }\over f},\quad {g_{l\nu ^\prime }g{ } _{j\nu }\over |V_{l\nu
 ^\prime }-V_{j\nu }|f}\right)  ,\qquad l\neq j\qquad \label{trts}
\end{equation}
and
\begin{equation}
R^\prime ,R^{\prime\prime}\gg {g_{j\nu ^\prime }g{ } _{j\nu }\over
|V_{j\nu ^\prime }-V_{j\nu }|f} . \label{trms}
\end{equation}
The scattering matrix (see (5.2) in \cite{YB98}) may then be
rewritten, using the present notation, as \begin{equation}
\eqalign{
\fl S^{\text{lin}}_{00}=\exp\left( -2\Lambda _{10}+2i\Lambda ^\prime
 \right)  \\
\fl S^{\text{lin}}_{j\nu ,0}=\sqrt{1-\exp\left( -2\lambda _{j\nu
 }\right) } \exp\left( -\Lambda _{j\nu }-\Lambda _{10}+2i\Lambda
 ^\prime \right)  \nonumber \\
\fl S^{\text{lin}}_{-j\nu ,0}=-\sqrt{1-\exp\left( -2\lambda _{j\nu
 }\right) } \exp\left( \Lambda _{j\nu }-\Lambda _{10}+\lambda _{j\nu
 }\right) ,\qquad S^{\text{lin}}_{-l\nu ^\prime ,-j\nu }=0 \nonumber\\
\fl S^{\text{lin}}_{l\nu ^\prime ,j\nu }=\sqrt{\left( 1-\exp\left(
-2\lambda _{l\nu ^\prime }\right) \right) \left( 1-\exp\left(
-2\lambda _{j\nu }\right) \right) }\exp\left( -\Lambda _{l\nu ^\prime
 }-\Lambda _{j\nu }+2i\Lambda ^\prime \right)  \nonumber\\
\fl S^{\text{lin}}_{-l\nu ^\prime ,j\nu }=-\sqrt{\left( 1-\exp\left(
-2\lambda _{l\nu ^\prime }\right) \right) \left( 1-\exp\left(
-2\lambda _{j\nu }\right) \right) } \\
\times \exp\left( \Lambda _{l\nu ^\prime }-\Lambda _{j\nu }+\lambda
 _{l\nu ^\prime }\right) \qquad \left( V_{l\nu ^\prime }>V_{j\nu
 }\right)  \nonumber\\
\fl S^{\text{lin}}_{-j\nu ,j\nu }=\exp\left( -\lambda _{j\nu }\right)
  \label{Sn}\\
\fl S^{\text{lin}}_{-l\nu ^\prime ,j\nu }=0\qquad \left( V_{l\nu
 ^\prime }<V_{j\nu }\right) . \nonumber
}
\end{equation}The remaining scattering matrix elements are
obtained by time-reversal symmetry, $S^{\text{lin}}_{k^\prime \nu
 ^\prime ,k\nu }=S^{\text{lin}}_{k\nu ,k^\prime \nu ^\prime }$. Here
 channels
$\pm j\nu $ correspond to the system in the state $|j\nu \rangle $ at
 $R\rightarrow \pm \infty $,
and\begin{equation}
\eqalign{
\lambda _{j\nu }=\pi {\mu g{ } ^{2}_{j\nu }\over p_{j\nu }f} ,\qquad
 p_{j\nu }=\sqrt{2\mu \left( E_{0}-V_{j\nu }\right) }\qquad \nonumber
 \\
\Lambda _{j\nu }=\sum\limits^{}_{\nu ^\prime >\nu }\lambda _{j\nu
 ^\prime }+\sum\limits^{n}_{j^\prime =j+1} \sum\limits^{d{ }
 _{j^\prime }}_{\nu ^\prime =1}\lambda _{j^\prime \nu ^\prime }
,\qquad \Lambda ^\prime =\sum\limits^{m}_{j=n+1} \sum\limits^{d{ }
 _{j}}_{\nu =1}|\lambda _{j\nu }|. \label{lam}
}
\end{equation} The elements of the scattering matrix (\ref{Sn}) have
the form of a product of LZ amplitudes. The counterintuitive
transitions are forbidden here, as can be seen from the last equality
of (\ref{Sn}).

The situation changes once condition (\ref{trms}) is removed;
i.e., if the quasi-degenerate states are close enough, given a
certain truncation range. We show here that, under appropriate
conditions, one can treat this case by starting from the approach
described in \cite{YB98} for the degenerate problem.

The orthogonal transformation
\begin{equation}
|j\kappa \rangle ^\prime =\sum\limits^{d{ } _{j}}_{\nu =1}A^{\left(
 j\right) }_{\kappa \nu }|j\nu \rangle ,\qquad a_{j\kappa }^\prime
 \left( R\right) =\sum\limits^{d{ } _{j}}_{\nu =1}A^{\left( j\right)
 }_{\kappa \nu }a_{j\nu }\left( R\right)  \label{TrBas}
\end{equation}
performed by the matrix
\begin{equation}
A^{\left( j\right) }_{0\nu }=g_{j\nu }/g_{j},\qquad g_{j}=\left(
 \sum\limits^{d{ } _{j}}_{\nu =1}g^{2}_{j\nu }\right) ^{1/2}
 \label{Aj}
\end{equation}
described in \cite{KF85,YB98},  leaves only one ($\kappa =0$) of the
 new
basis states in the $j$-th group coupled to the sloped potential
channel. Unlike the strictly-degenerate case considered in
\cite{YB98}, in the quasi-degenerate case this transformation leads
to the non-diagonal potential matrix
\begin{equation}
V^{\left( j\right) }_{\kappa ^\prime \kappa }= \sum\limits^{d{ }
 _{j}}_{\nu =1}A^{\left( j\right) }_{\kappa ^\prime \nu }V_{j\nu
 }A^{\left( j\right) }_{\kappa \nu }.\qquad \label{PotMat}
\end{equation}
The transformed close-coupling coefficients $a_{j\kappa }^\prime
 \left( R\right) $ obey the
following equations\begin{eqnarray}
- {1\over 2\mu } {\partial ^{2}a_{j\kappa }^\prime \over \partial R{
 } ^{2}}+\sum\limits^{d_{j}-1}_{\kappa ^\prime =0}V^{\left( j\right)
 }_{\kappa \kappa ^\prime }a_{j\kappa ^\prime }^\prime
=E_{0}a_{j\kappa }^\prime \qquad \left( \kappa \neq 0\right)
 \label{TrEaNi}\\
- {1\over 2\mu } {\partial ^{2}a_{j0}^\prime \over \partial R{ }
 ^{2}}+\sum\limits^{d_{j}-1}_{\kappa =0}V^{\left( j\right) }_{0\kappa
 }a_{j\kappa }^\prime +g_{j}b=E_{0}a_{j0}^\prime \qquad
 \label{TrEaI}\\
- {1\over 2\mu } {\partial ^{2}b\over \partial R{ } ^{2}} - f R b
+\sum\limits^{m}_{j=1}g_{j}a_{j0}^\prime =0. \label{TrEb}
\end{eqnarray}Thus, the non-interacting channels ($\kappa \neq 0$)
 will be
coupled with other channels in this quasi-degenerate group by the
non-diagonal elements of the matrix (\ref{PotMat}). In the case of
strict degeneracy, $V_{j\nu }$  is $\nu $-independent and the matrix
(\ref{PotMat}) is then diagonal, as required in \cite{YB98}.

\subsection{Perturbation theory}

Let us solve the equations (\ref{TrEaNi})-(\ref{TrEb}) under
conditions in which the non-diagonal elements of the potential
matrix (\ref{PotMat}) may be considered as a small perturbation. The
orthogonality of the matrix $A^{\left( j\right) }_{\kappa \nu }$
 allows us to evaluate the
magnitude of these elements in terms of characteristic width of the
quasi-degenerate group, $\Delta V_{j}$, defined as
\begin{equation}
\sum\limits^{d_{j}-1}_{\kappa ,\kappa ^\prime =0}\left( V^{\left(
 j\right) }_{\kappa \kappa ^\prime }-V^{\left( j\right) }_{00}\delta
 _{\kappa \kappa ^\prime }\right) ^{2}=\sum\limits^{d{ } _{j}}_{\nu
=1}\left( V_{j\nu }-V^{\left( j\right) }_{00}\right) ^{2}=d_{j}\Delta
 V^{2}_{j} .
\end{equation}
Thus, a small $\Delta V_{j}$  means a weak perturbation.

The unperturbed equations are similar to those used in the
degenerate case (see \cite{YB98}). The unperturbed equations
(\ref{TrEaNi}) are uncoupled. Therefore the curve-crossing problem
has the following unit-flux normalized plane-wave solutions
\begin{equation}
\varphi _{\pm j\kappa }=\left( \mu /p_{j\kappa }^\prime \right) ^{1
/2}\exp\left( \mp ip_{j\kappa }^\prime R\right) |j\kappa \rangle
 ^\prime ,\qquad \kappa \neq 0 \label{unpni}
\end{equation}
where the signs $\pm $ denote the location of the source of the
incoming wave $\left( \pm \infty \right) $, and
\begin{equation}
p_{j\kappa }^\prime =\sqrt{2\mu \left( E_{0}-V^{\left( j\right)
 }_{\kappa \kappa }\right) }.
\end{equation}
The unperturbed equations (\ref{TrEaI}) and (\ref{TrEb})
describe the non-degenerate linear curve-crossing problem considered
in \cite{YB98}. Thus the remaining solutions of the unperturbed
problem may be expressed in terms of the fundamental solutions
 $a_{j}\left( R\right) $
and $b\left( R\right) $, introduced in \cite{YB98}. The solution,
 containing a
unit-flax incoming wave in the state $|0\rangle $, and an outgoing
 wave
containing all other coupled states, has the form
\begin{equation}
\varphi _{0}=e^{-\Lambda _0}\left\lbrack b\left( R\right) |0\rangle
+\sum\limits^{m}_{j=1}a_{j}\left( R\right) |j0\rangle ^\prime
 \right\rbrack \qquad \label{unp0}
\end{equation}
where $a_{j}\left( R\right) $ and $b\left( R\right) $ are defined by
 equations (4.1)-(4.3) in
\cite{YB98}, with $p_{j}=p_{j0}^\prime $ and $s_{\pm l}=1$ for all
 $l$. Here
\begin{equation}
\Lambda _{j}=\sum\limits^{n}_{j^\prime =j+1}\lambda _{j^\prime }
,\qquad \lambda _{j}=\pi {\mu g{ } ^{2}_{j}\over p_{j0}^\prime f}.
\end{equation}
The solution containing unit-flux incoming waves in the states
$|j\kappa \rangle ^\prime $ may be constructed using other choices
 for $s_{\pm l}$\footnote{Values
of $s_{\pm l}$  different from 1 will be marked below as additional
arguments of $a_{j}$  and $b$.}. So, the solutions representing  waves
incoming from the negative $R$ direction have the form\begin{eqnarray}
\fl \varphi _{-j0}=\left( 2\sinh\lambda _{j}\right) ^{-1/2}\exp\left(
 -\lambda _{j}/2+\Lambda _{j}-2i\Lambda ^\prime \right)
 \{\left\lbrack b\left( R,s_{+j}=-1\right) -b\left( R\right)
 \right\rbrack |0\rangle  \nonumber\\
+\sum\limits^{m}_{l=1}\left\lbrack a_{l}\left( R,s_{+j}=-1\right)
-a_{l}\left( R\right) \right\rbrack |l0\rangle ^\prime \} .
 \label{unpmj0}
\end{eqnarray}The remaining solutions, representing waves
incoming from the positive $R$ direction, have the form
\begin{eqnarray}
\fl \varphi _{+j0}=\left( 2\sinh\lambda _{j}\right) ^{-1/2}\exp\left(
 \lambda _{j}/2-\Lambda _{j}\right) \{\left\lbrack b\left( R,s_{-j}=
-1\right) -e^{-2\lambda _j}b\left( R\right) \right\rbrack |0\rangle
 \nonumber
\\
+\sum\limits^{m}_{l=1}\left\lbrack a_{l}\left( R,s_{-j}=-1\right)
-e^{-2\lambda _j}a_{l}\left( R\right) \right\rbrack |l0\rangle
 ^\prime \} . \label{unppj0}
\end{eqnarray}
The evaluation of the perturbation matrix elements connecting
the unperturbed wavefunctions (\ref{unp0})-(\ref{unppj0}) and
(\ref{unpni}) includes an integration over $R$ of the products of the
exponential function from (\ref{unpni}) with $a_{j}\left( R\right) $.
 (One does not
have to evaluate similar integrals with $b\left( R\right) $, since
 the states $|0\rangle $
and $|j\kappa \rangle ^\prime $, with $\kappa \neq 0$, are not
 coupled.) Using the contour-integral
representation (3.2) in \cite{YB98}, the integral may be transformed
into the following form,
\begin{eqnarray}
\fl \int\limits^{R^{\prime\prime}}_{-R^\prime }dR \exp\left( \mp
 ip_{j\kappa }^\prime R\right) a_{j}\left( R\right)  \nonumber
\\
=-i\int\limits^{}_{C}dp \tilde{a}_{j}\left( p\right) {\exp\left(
 i\left( p\pm p_{j\kappa }^\prime \right) R^{\prime\prime}\right)
-\exp\left( -i\left( p\pm p_{j\kappa }^\prime \right) R^\prime \right
) \over p\pm p_{j\kappa }^\prime } .
\end{eqnarray}
If the conditions (\ref{trts}) are satisfied, the asymptotic
expansion of the integral may be evaluated in the manner used in
\cite{YB98} for the evaluation of $a_{j}\left( R\right) $. (The
 integration contour
should enclose all the poles $\pm p_{j\kappa }^\prime $  for each $j$
 simultaneously.) As a
result, the matrix elements are expressible as\begin{equation}
\eqalign{
\fl \langle \varphi _{\pm j\kappa }|V^{\left( j\right) }|\varphi
 _{0}\rangle =S_{j0}\xi ^{\pm }_{j\kappa }\left(
 R^{\prime\prime}\right) -S_{-j0}\xi ^{\mp *}_{j\kappa }\left(
-R^\prime \right) \qquad \nonumber \\
\fl \langle \varphi _{\pm j\kappa }|V^{\left( j\right) }|\varphi
 _{j^\prime 0}\rangle =\delta _{jj^\prime }\xi ^{\mp *}_{j\kappa
 }\left( R^{\prime\prime}\right) +S_{jj^\prime }\xi ^{\pm }_{j\kappa
 }\left( R^{\prime\prime}\right) -S_{-jj^\prime }\xi ^{\mp
*}_{j\kappa }\left( -R^\prime \right)  \label{inme} \\
\fl \langle \varphi _{\pm j\kappa }|V^{\left( j\right) }|\varphi _{
-j^\prime 0}\rangle =-\delta _{jj^\prime }\xi ^{\pm }_{j\kappa }\left
( -R^\prime \right) +S_{j-j^\prime }\xi ^{\pm }_{j\kappa }\left(
 R^{\prime\prime}\right)  . \nonumber
}
\end{equation} Here \begin{equation}
\eqalign{
\fl \xi ^{+}_{j\kappa }\left( R\right) =-i\left( \mu /p_{j\kappa
 }^\prime \right) ^{1/2}V^{\left( j\right) }_{\kappa 0}{\alpha
 _{j}\left( R\right) \over p_{j0}^\prime +p_{j\kappa }^\prime }
 \exp\left( ip_{j\kappa }^\prime R\right) \qquad \label{csipm} \\
\fl \xi ^{-}_{j\kappa }\left( R\right) =i\left( \mu /p_{j\kappa
 }^\prime \right) ^{1/2}V^{\left( j\right) }_{\kappa 0}\left\lbrack
 {1\over p_{j0}^\prime +p_{j\kappa }^\prime }+{\pi R\over \lambda
 _{j}+i\pi }\right\rbrack  \alpha _{j}\left( R\right) \exp\left(
-ip_{j\kappa }^\prime R\right)  \nonumber
}
\end{equation} in which $\alpha _{j}\left( R\right) $ are a set of
 waves of unit-flux
normalization appearing in the asymptotic solution, as defined by
(4.9) of \cite{YB98}. The amplitudes $S_{kk^\prime }$  are the
 elements of the
scattering matrix for the non-degenerate case defined by (5.2) of
\cite{YB98}. They are also obtained from (\ref{Sn}) by setting $d_{j}
=1$
for all $j$ and omitting the subscript $\nu $, i.e.,
\begin{equation}
\fl S_{kk^\prime }=S^{\text{lin}}_{k1,k^\prime 1}\qquad \left( k
,k^\prime \neq 0\right) ,\qquad S_{k0}=S^{\text{lin}}_{k1,0},\qquad
 S_{0k}=S^{\text{lin}}_{0,k1}. \label{Snd}
\end{equation}
The perturbation matrix elements between the states
(\ref{unpni}) have the form
\begin{eqnarray}
\fl \langle \varphi _{\sigma ^\prime j\kappa ^\prime }|V^{\left(
 j\right) }|\varphi _{\sigma j\kappa }\rangle =i\mu \left( p_{j\kappa
 }^\prime p_{j\kappa ^\prime }^\prime \right) ^{-1/2}V^{\left(
 j\right) }_{\kappa ^\prime \kappa } \nonumber
\\
\fl \times {\exp\left( -i\left( \sigma p_{j\kappa }^\prime -\sigma
 ^\prime p_{j\kappa ^\prime }^\prime \right) R^{\prime\prime}\right)
-\exp\left( i\left( \sigma p_{j\kappa }^\prime -\sigma ^\prime
 p_{j\kappa ^\prime }^\prime \right) R^\prime \right) \over \sigma
 p_{j\kappa }^\prime -\sigma ^\prime p_{j\kappa ^\prime }^\prime
 }\qquad \left( \sigma ,\sigma ^\prime =\pm \right)  . \label{nime}
\end{eqnarray}
The transitions between the unperturbed states are negligible if
the matrix elements (\ref{inme}) and (\ref{nime}) are small compared
to a unit. Since $S_{kk^\prime }\le 1$, the matrix elements
(\ref{inme}) are small
if the functions (\ref{csipm}) are small. This imposes the following
restrictions on $R^\prime $ and $R^{\prime\prime}$,
\begin{equation}
R^\prime ,R^{\prime\prime}\ll {1\over \Delta V{ } _{j}}\sqrt{{E_{0}
-V{ } _{j\nu }\over \mu }} + {g{ } ^{2}_{j}\over f\Delta V{ }
 _{j}}\qquad \left( j\le m\right)  \label{condr}
\end{equation}
which is the opposite of the condition of applicability
(\ref{trms})  of the solution for the non-degenerate case. The full
condition of negligibility of the perturbation effect may be written
as a restriction on the characteristic width of quasi-degenerate
groups
\begin{equation}
\Delta V_{j}\ll \min \left( 4\left( E_{0}-V_{j\nu }\right) ,\quad
 {p_{j}\left( 1+\lambda _{j}/\pi \right) \over \mu R^\prime } ,\quad
 {p_{j}\left( 1+\lambda _{j}/\pi \right) \over \mu
 R^{\prime\prime}}\right)  \label{condv}
\end{equation}
recalling that\begin{equation}
\eqalign{
{p_{j}\left( 1+\lambda _{j}/\pi \right) \over \mu R^\prime }={1\over
 R^\prime }\sqrt{{E_{0}-V{ } _{j\nu }\over \mu }} + {g{ }
 ^{2}_{j}\over |V_{0}\left( -R^\prime \right) -E_{0}|} \nonumber \\
{p_{j}\left( 1+\lambda _{j}/\pi \right) \over \mu R^{\prime\prime}}
={1\over R^{\prime\prime}}\sqrt{{E_{0}-V{ } _{j\nu }\over \mu }} +
 {g{ } ^{2}_{j}\over |V_{0}\left( R^{\prime\prime}\right) -E_{0}|} .
}
\end{equation} Thus, if these conditions are obeyed, the
transitions in a truncated quasi-degenerate system may be described
by using the scattering matrix for the degenerate system; i.e., (5.6)
and (5.7) in \cite{YB98},
\begin{eqnarray}
S_{k^\prime \nu ^\prime ,k\nu }={g_{|k^\prime |\nu ^\prime }g{ } _{|k
|\nu }\over g_{|k^\prime |}g{ } _{|k|}}S_{k^\prime k}+\left( \delta
 _{\nu \nu ^\prime }-{g_{|k|\nu ^\prime }g{ } _{|k|\nu }\over g{ }
 ^{2}_{|k|}}\right) \delta _{-kk^\prime } \label{Sg}
\\
S_{k\nu ,0}={g{ } _{|k|\nu }\over g{ } _{|k|}}S_{k0},\qquad S_{0,k\nu
 }={g{ } _{|k|\nu }\over g{ } _{|k|}}S_{0k}\qquad \label{Sg0}
\end{eqnarray}
where $S_{k^\prime k}$  are defined by (\ref{Snd}) or (5.2) of
 \cite{YB98}.
The scattering matrix (\ref{Sg}) cannot be represented in the
semiclassical form as a product of LZ amplitudes, and allows for
counterintuitive transitions ($k^\prime =-k<0$, $\nu ^\prime <\nu $).

\section{Piecewise-linear problem}

\subsection{Transitions in the external regions}

In the previous section it was shown that transitions in the
quasi-degenerate system confined to a finite vicinity of the crossing
points, defined by the conditions (\ref{condr}), may be described by
the scattering matrix for the degenerate system (\ref{Sg}) and
(\ref{Sg0}). However, transitions between the quasi-degenerate states
do not stop at the edges of this vicinity. Transitions in the
external regions beyond this vicinity ultimately lead to the
scattering matrix for the non-degenerate system (\ref{Sn}). Let us
introduce orthogonal matrices $B^{\left( k\right) }_{\kappa \nu }$
 describing the transitions in the
external regions ($R>R^{\prime\prime}$ for $k>0$ and $R<-R^\prime $
 for $k<0$) between the
asymptotic states $||k|\nu \rangle $ at infinity and the states $||k
|\kappa \rangle ^\prime $ at the
edges of the internal region. These matrices are diagonal with
respect to transitions between states of different quasi-degenerate
groups since these are fully accomplished within the internal region
when the conditions (\ref{trts}) are obeyed. Thus, the scattering
matrix (\ref{Sn}) for the quasi-degenerate system in the infinite
range can be approximately expressed in the form
\begin{eqnarray}
\fl S^{\text{lin}}_{k^\prime \nu ^\prime ,k\nu }\approx B^{\left(
 k^\prime \right) }_{0\nu ^\prime }B^{\left( k\right) }_{0\nu
 }S_{k^\prime k}+\sum\limits^{}_{\kappa \neq 0}B^{\left( k^\prime
 \right) }_{\kappa \nu ^\prime }B^{\left( k\right) }_{\kappa \nu
 }\delta _{-kk^\prime }\qquad \left( k,k^\prime \neq 0\right) \qquad
 \label{Slin}
\\
\fl S^{\text{lin}}_{k\nu ,0}=S^{\text{lin}}_{0,k\nu }\approx B^{\left
( k\right) }_{0\nu }S_{k0}.\qquad \label{Slin0}
\end{eqnarray}
Here use was made of properties (5.3)-(5.5) of \cite{YB98}
concerning the scattering matrix for the degenerate states in the
transformed basis.

In the limit of strict degeneracy, where the matrix elements
(\ref{inme}) and (\ref{nime})  vanish, and transitions between the
states of the transformed basis cease to exist, $B^{\left( k\right)
 }_{\kappa \nu }=A^{\left( |k|\right) }_{\kappa \nu }$, and
(\ref{Slin}) and (\ref{Slin0}) are reduced to (\ref{Sg}) and
(\ref{Sg0}), respectively, defining the scattering matrix in the
degenerate case.

The substitution of (\ref{Sn}) for $S^{\text{lin}}_{k\nu ,0}$  and
(\ref{Snd}) for
$S_{k0}$  allows us to obtain the following exact expressions for
$B^{\left( k\right) }_{0\nu }$,\begin{equation}
\eqalign{
B^{\left( j\right) }_{0\nu }=\left\lbrack {1-\exp\left( -2\lambda
 _{j\nu }\right) \over 1-\exp\left( -2\lambda _{j}\right)
 }\right\rbrack ^{1/2}\exp\left( \Lambda _{j}-\Lambda _{j\nu }\right)
  \label{B0} \\
B^{\left( -j\right) }_{0\nu }=\left\lbrack {1-\exp\left( -2\lambda
 _{j\nu }\right) \over 1-\exp\left( -2\lambda _{j}\right)
 }\right\rbrack ^{1/2}\exp\left( \Lambda _{j\nu }-\Lambda _{j}
+\lambda _{j\nu }-\lambda _{j}\right)  .
}
\end{equation} These expressions also obey (\ref{Slin}) exactly
for $k\neq k^\prime $  being independent of $B^{\left( k\right)
 }_{\kappa \nu }$  with $\kappa \neq 0$.

Hereafter we shall consider only the case in which each
quasi-degenerate group consists of two states only ($d_{j}=2$,
$\nu =1,2$, and $\kappa =0,1$). In this case the remaining elements of
$B^{\left( k\right) }_{\kappa \nu }$  are defined by the
 orthogonality of this matrix,
resulting in
\begin{equation}
B^{\left( \pm j\right) }_{1\nu }=\left( -1\right) ^{\nu -1}\sigma
 _{\pm }B^{\left( \pm j\right) }_{03-\nu }\qquad \label{B1}
\end{equation}
where $\sigma _{\pm }$  may be chosen as either +1 or -1.

The matrix $B^{\left( k\right) }_{\kappa \nu }$  obtained in this
 manner obeys the equations
(\ref{Slin}) only approximately. By choosing $\sigma _{+}=\sigma _{
-}$,  the residuals
become smaller than
\begin{equation}
{\lambda _{j}-\lambda _{j1}-\lambda { } _{j2}\over \lambda { }
 _{j}}\approx {\Delta V{ } _{j}\over 2\left( E_{0}-V^{\left( j\right)
 }_{00}\right) } \label{DS}
\end{equation}
and may be neglected whenever the first criterion in
(\ref{condv}) is obeyed.

One may expect the inaccuracy of the representation
(\ref{Slin0}), (\ref{Slin}) to be of the same order as the matrix
elements (\ref{inme}) and (\ref{nime}). However, the inaccuracy is
independent on $R^\prime $ and $R^{\prime\prime}$. This means that
 the corresponding errors in
the transition amplitudes in the central and external regions cancel
each other in this case of a linear sloped potential. Therefore, the
elements $B^{\left( k\right) }_{\kappa \nu }$  provide an estimate of
 the transition amplitudes in
the external regions, to the same accuracy as that provided by the
scattering matrix of the degenerate case for the transition
amplitudes in the central region. The amount of inaccuracy may be
estimated by  the matrix elements (\ref{inme}) and (\ref{nime}).

The orthogonality conditions are sufficient to determine the
matrix $B^{\left( \pm j\right) }_{\kappa \nu }$  for $d_{j}=2$ only,
 since only in a two-dimensional space a
given vector (the row $B^{\left( \pm j\right) }_{0\nu }$) has only
 one unit vector orthogonal to
it (up to a sign $\sigma _{\pm }$). The relative signs of $B^{\left(
+j\right) }_{1\nu }$   and $B^{\left( -j\right) }_{1\nu }$  are
chosen so as to produce minimal residuals on substitution to
(\ref{Slin}), the absolute signs being insignificant. In the case of
a $d_{j}$-dimensional space with $d_{j}>2$ there are $d_{j}-1$
 mutually orthogonal
vectors which are orthogonal to the given vector (the row $B^{\left(
 \pm j\right) }_{0\nu }$). The
matrix $B^{\left( \pm j\right) }_{\kappa \nu }$  may also be
 considered as consisting of $d_{j}$  mutually
orthogonal vectors-columns with fixed components $B^{\left( \pm
 j\right) }_{0\nu }$. These vectors
are defined up to a rotation (about the $\kappa =0$ unit vector),
characterized by $d_{j}-2$ arbitrary angles, and (\ref{Slin}) may then
define only a relative rotation of $B^{\left( +j\right) }_{\kappa \nu
 }$  and $B^{\left( -j\right) }_{\kappa \nu }$.

\subsection{Total scattering matrix}

As was shown in the previous subsection, the scattering matrix
for the quasi-degenerate linear problem may be approximately
represented as a product of the scattering matrix for a degenerate
problem, describing transitions in a finite vicinity of the crossing
points, and the matrices $B^{\left( k\right) }_{\kappa \nu }$,
 describing transitions in the external
wings. This fact allows us to consider a piecewise-linear problem
(see figure 1), in which the sloped potential consists of three
segments of varying slopes ($-f^\prime $ at $R<-R^\prime $, $-f$ at $
-R^\prime <R<R^{\prime\prime}$, and $-f^{\prime\prime}$ at
$R>R^{\prime\prime}$). If $R^\prime $, $R^{\prime\prime}$, and
 $\Delta V_{j}$  obey the conditions (\ref{condv}), we can
associate the transitions at $R<-R^\prime $ with the matrix $B^{\left
( -j\right) }_{\kappa \nu }\left( f^\prime \right) $, at
$-R^\prime <R<R^{\prime\prime}$ with $S_{kk^\prime }\left( f\right) $
, and at $R>R^{\prime\prime}$ with $B^{\left( j\right) }_{\kappa \nu
 }\left( f^{\prime\prime}\right) $. The $f$ arguments
refer to the forces with which these matrices should be evaluated.
The total scattering matrix can then be written as \begin{equation}
\eqalign{
\fl S^{pl}_{k^\prime \nu ^\prime ,k\nu }\approx B^{\left( k^\prime
 \right) }_{0\nu ^\prime }B^{\left( k\right) }_{0\nu }S_{k^\prime
 k}\left( f\right) +\sum\limits^{}_{\kappa \neq 0}B^{\left( k^\prime
 \right) }_{\kappa \nu ^\prime }B^{\left( k\right) }_{\kappa \nu
 }\delta _{-kk^\prime }\qquad \left( k,k^\prime \neq 0\right)  \\
\fl S^{pl}_{k\nu ,0}\approx B^{\left( k\right) }_{0\nu }S_{k0}\left(
 f\right) ,\qquad S^{pl}_{0,k\nu }\approx B^{\left( k\right) }_{0\nu
 }S_{0k}\left( f\right)  \label{Spl}
}
\end{equation} where $B^{\left( k\right) }_{\kappa \nu }$  is taken
 as $B^{\left( k\right) }_{\kappa \nu }\left( f^{\prime\prime}\right)
 $ if $k>0$ and as
$B^{\left( k\right) }_{\kappa \nu }\left( f^\prime \right) $ if $k<0$
. The same sign $\sigma _{\pm }$  has been chosen for $B^{\left(
-j\right) }_{\kappa \nu }\left( f^\prime \right) $ and
$B^{\left( +j\right) }_{\kappa \nu }\left( f^{\prime\prime}\right) $
 in order to maintain continuity at $f^\prime =f$ and
 $f^{\prime\prime}=f$. (The
angles describing the arbitrary rotations about the direction of the
interacting state if $d_{j}>2$, being continuous parameters, may not
 be
completely determined in this manner.) The elements of the scattering
matrix (\ref{Spl}) cannot be represented as a product of LZ
amplitudes.

It is interesting to consider in more detail the
elements $S^{pl}_{-j\nu ^\prime ,j\nu }$  ($j>0$) describing
 transmission within the
same quasi-degenerate group. Substituting (\ref{B0}) and
(\ref{B1}), as well as (\ref{Snd}), into (\ref{Spl}) one
obtains
\begin{eqnarray}
\fl S^{pl}_{-j\nu ^\prime ,j\nu }=\left\lbrack 1-\exp\left( -2\lambda
 _{j}^\prime \right) \right\rbrack ^{-1/2}\left\lbrack 1-\exp\left(
-2\lambda _{j}^{\prime\prime}\right) \right\rbrack ^{-1/2} \nonumber
\\
\fl \times \biggl\lbrack \sqrt{\left( 1-\exp\left( -2\lambda _{j\nu
 ^\prime }^\prime \right) \right) \left( 1-\exp\left( -2\lambda
 _{j\nu }^{\prime\prime}\right) \right) }\exp\left( -\lambda _{j}
-\lambda _{j1}^\prime \delta _{\nu ^\prime 2}-\lambda
 _{j2}^{\prime\prime}\delta _{\nu 1}\right)  \nonumber
\\
\fl +\left( -1\right) ^{\nu -\nu ^\prime }\sqrt{\left( 1-\exp\left(
-2\lambda _{j3-\nu ^\prime }^\prime \right) \right) \left( 1
-\exp\left( -2\lambda _{j3-\nu }^{\prime\prime}\right) \right) }
 \nonumber
\\
\times \exp\left( -\lambda _{j1}^\prime \delta _{\nu ^\prime 1}
-\lambda _{j2}^{\prime\prime}\delta _{\nu 2}\right)  \biggr\rbrack
 \label{Tpl}
\end{eqnarray}
where $\lambda _{j\nu }^\prime $  and $\lambda _{j\nu
 }^{\prime\prime}$ are defined by (\ref{lam}) with $f$
replaced by $f^\prime $  or $f^{\prime\prime}$, respectively. The
 counterintuitive
transitions then correspond to the matrix elements
\begin{eqnarray}
\fl S^{pl}_{-j1,j2}=\left\lbrack 1-\exp\left( -2\lambda _{j}^\prime
 \right) \right\rbrack ^{-1/2}\left\lbrack 1-\exp\left( -2\lambda
 _{j}^{\prime\prime}\right) \right\rbrack ^{-1/2} \nonumber
\\
\times \biggl\lbrack \sqrt{\left( 1-\exp\left( -2\lambda _{j1}^\prime
 \right) \right) \left( 1-\exp\left( -2\lambda
 _{j2}^{\prime\prime}\right) \right) }\exp\left( -\lambda _{j}\right)
  \nonumber
\\
-\sqrt{\left( 1-\exp\left( -2\lambda _{j2}^\prime \right) \right)
 \left( 1-\exp\left( -2\lambda _{j1}^{\prime\prime}\right) \right)
 }\exp\left( -\lambda _{j1}^\prime -\lambda
 _{j2}^{\prime\prime}\right)  \biggr\rbrack  . \label{Tplci}
\end{eqnarray}
In the limit $f^\prime =f^{\prime\prime}=f$ these amplitudes become
 smaller than
(\ref{DS}), which serves as a measure of the inaccuracy of this
approximation.

Of special interest is the case in which the potential $V_{0}$  in
one of the external wings is horizontal ($f^\prime =0$ or
 $f^{\prime\prime}=0$). In this case
the present approach is formally inapplicable, since the finite gap
between $V_{0}$  and the other potentials does not allow to neglect
 the
interaction even at infinity and to set the asymptotic boundary
conditions with incoming flux in one channel only. This case may be
treated by assuming that the interaction constants $g_{j\nu }$  are
 gradually
turned off towards infinity. This assumption is in agreement with
real physical situations. For example, the laser beam inducing the
coupling of atomic states, has a finite width, very large in terms of
atomic dimensions.

An adiabatically-slow turning on of the interaction makes the
system stay in the same adiabatic state. These states are obtained by
diagonalization of the potential matrix including the interactions.
If the width of the quasi-degenerate group satisfies the condition
(\ref{condv}) with $p_{j}=0$, and the potential $V_{0}$  is far
 distanced from
other potentials [the conditions (\ref{trts}) being  sufficient for
this case], the adiabatic states are nothing else but the states
$|j\kappa \rangle ^\prime $  introduced in (\ref{TrBas}). In the case
 of $d_{j}=2$, the
adiabatic energy of the interacting channel state $|j0\rangle ^\prime
 $ and $V_{0}$  lie
in the opposite sides of the adiabatic energy of the non-interacting
channel state $|j1\rangle ^\prime $. Since the adiabatic potentials
 do not cross each
other the state $|j0\rangle ^\prime $  corresponds adiabatically to $
|j2\rangle $ as $R\rightarrow \infty $ and
to $|j1\rangle $ as $R\rightarrow -\infty $. This fact is also known
 in the theory of ``dark
states'' (see \cite{DarkStates} and references therein). A more
detailed analysis yields
\begin{equation}
B^{\left( +j\right) }_{\kappa \nu }=1-\delta _{\kappa ,\nu -1},\qquad
 B^{\left( -j\right) }_{\kappa \nu }=\left( -1\right) ^{\nu }\delta
 _{\kappa ,\nu -1}\qquad \label{Bflat}
\end{equation}
which coincide with the limiting values of (\ref{B0}) and
(\ref{B1}) as $f^\prime \rightarrow 0$ or
 $f^{\prime\prime}\rightarrow 0$. Thus, the results of the present
 theory
are applicable to this case as well.

\section{Comparison with numerical results}

In order to test the approximations used in the present
theory, the scattering matrix was evaluated by using the
analytical theory provided here, and also calculated numerically,
by using the invariant imbedding method \cite{imb} to solve the
associated close-coupling equations, for a specific model. This
model involves only two horizontal potentials ($d_{1}=2$) forming one
quasi-degenerate group ($n=m=1$).  The parameters of the model were
chosen so as to simulate an optical collision of metastable Xe
atoms (see \cite{YB97}). Here $\mu =66$ AMU (1 AMU = $1.6605\times
 10^{-27}$  Kg),
collision energy $E=10^{-9}$  au (1 au = $4.3597\times 10^{-18}$  J),
 and
$f=2.17\times 10^{-10}$  au/a$_{0}$  (a$_{0}= 0.0529177$ nm).  The
 coupling constants,
which are dependent on the laser intensity $I$, were taken as
$g_{11}=9.6\times 10^{-9}\lbrack I$(W/cm$^{2}$)$\rbrack ^{1/2}$  au
 and $g_{12}=5.6\times 10^{-9}\lbrack I$(W/cm$^{2}$)$\rbrack ^{1/2}$
 au.
The first criterion in (\ref{trts}) requires a large value of
$R^\prime =R^{\prime\prime}=6\times 10^{3}$  a$_{0}$. This value,
 dictated by the small value of the
kinetic energy, is larger than the range one would normally
associate with the shielding process simulated by this model.

The results for the truncated linear problem are presented in
figure 2, using three different values of the potential gap. The
calculations show that for the small gap $V_{12}-V_{11}=6.7\times
 10^{-12}$  au
(figure $2a$) the expressions (\ref{Sg}) and (\ref{Sg0}) for the
scattering matrix in the degenerate case are in good agreement with
the numerical results. For the large gap $V_{12}-V_{11}=5\times 10^{
-10}$  au (figure
$2c$) the agreement is better with the expressions (\ref{Sn}) for the
scattering matrix in the non-degenerate case. The numerical results
for the intermediate gap $V_{12}-V_{11}=6.7\times 10^{-11}$  au
(figure 2b)  lie
between the predictions of the two models. The latter case
corresponds to the actual gap between the energies at the crossing
points in which the $s$ and $d$ partial-wave potentials of the lower
(metastable) state of Xe cross the $p$ partial-wave potential of the
excited state. At high intensities, however, the numerical results
tend to the predictions of the quasi-degenerate model (as discussed
in section 5 below).

The numerical calculations for the piecewise-linear model are
somewhat more tedious, as the gap sizes used here require a very wide
integration range, reaching near-macroscopic dimensions. We have
conducted calculations for the case in which the two wings are flat
($f^\prime =f^{\prime\prime}=0$), keeping all other parameters the
 same as in the truncated
model discussed above. Figure 3 shows two transition elements (the
intuitive one above, the counterintuitive below), demonstrating
excellent agreement between the calculations and the analytical
results using (\ref{Spl}) and (\ref{Bflat}).

\section{Discussion}

We have analytically and numerically solved model problems that
are modifications of the Demkov-Osherov model of a sloped linear
potential curve crossing a set of horizontal ones.  Two types of
modifications were considered: (a) truncation, in which the boundary
conditions are determined at the ends of a finite interval, and (b)
modification of the sloped potential into a piecewise-linear form. The
modified problems can be treated by using the quasi-degeneracy
approximation, which is valid when the criteria (\ref{trts}) and
(\ref{condv}) are obeyed. This approximation means that the results of
the degenerate model discussed in \cite{YB98} should be used. This
model allows for counterintuitive transitions. The opposite happens
when criteria (\ref{trms}) are met. In this case, in which the
transition range lies within the range of the finite segment of the
sloped potential, the results of the non-degenerate (i.e., the
 original
Demkov-Osherov) model apply, in which case counterintuitive
 transitions
are forbidden. It follows from the present analysis that
counterintuitive transitions are generally quite common in situations
involving a sloped potential crossing several horizontal ones. In the
unmodified problem (dealt with by the Demkov--Osherov model),
contributions coming from different parts of the transition region
cancel each other, and lead to the disappearance of the
counterintuitive transitions. Such a compensation does not take place
anymore when the conditions of quasi-degeneracy (\ref{condv}) are
obeyed.

The criteria (\ref{condv}) allow for an interpretation that
stems from the viewpoint of the uncertainty principle. Let us denote
as $\Delta p_{j}=\mu \Delta V_{j}/p_{j}$  the characteristic
 difference of momenta in the
quasi-degenerate group for a given total energy, and as $t^\prime
=\mu R^\prime /p_{j}$
and $t^{\prime\prime}=\mu R^{\prime\prime}/p_{j}$  the characteristic
 times of travelling from $R^{\prime\prime}$  to 0
and from 0 to $R^\prime $, respectively. Using this notation, the
 criteria
(\ref{condv}) may be written in one of the following forms,
\begin{equation}
\eqalign{
\Delta p_{j}\max\left( R^\prime ,R^{\prime\prime}\right) \ll \hbar
 \left( 1+\lambda _{j}/\pi \right)  \\
\Delta V_{j}\max\left( t^\prime ,t^{\prime\prime}\right) \ll \hbar
 \left( 1+\lambda _{j}/\pi \right)  .
}
\end{equation} The first form means that the momenta in the
quasi-degenerate states are indistinguishable at the given coordinate
interval. The second one means that the potential energies of the
quasi-degenerate states are indistinguishable for the given
travelling time. The factor $\left( 1+\lambda _{j}/\pi \right) $
 describes a broading of the
uncertainty as the coupling increases. As one may see from figure 2,
the higher the intensity becomes, the larger is the value of $\Delta
 V_{j}$
applicable in the quasi-degenerate approximation.

The expansion of the applicability region in
the quasi-degenerate model as the coupling
constants increase leads to an interesting property
of the transmission amplitudes, that may be
interpreted as a stabilization effect. Let us
consider, for example, a case in which
$g_{j1}=g_{j2}=\ldots  =g_{jd_{j}}=d^{-1/2}_{j}g_{j}$. As long as the
 $g_{j\nu }$  are
small, the criteria (\ref{trms}) are obeyed, and
the system should be considered as a non-degenerate
one. The amplitude of elastic transmission in the
state $|j\nu \rangle $ is $S_{-j\nu ,j\nu }=\exp\left( -\lambda
 _{j\nu }\right) $ (see (\ref{Sn})).
This amplitude decreases exponentially as the
coupling constant increases. Upon further
increasing  $g_{j\nu }$  conditions (\ref{trms}) are
violated, but conditions (\ref{condv}) for the
applicability of the quasi-degeneracy approximation
are validated. The transmission amplitude
$S_{-j\nu ,j\nu }=1-\left\lbrack 1-\exp\left( \lambda _{j}\right)
 \right\rbrack /d_{j}$  (see (\ref{Sg})) is close
to unity if $d_{j}$  is large. Moreover, the higher the
coupling constants, the more states may be bunched
into the quasi-degenerate group, i.e., $d_{j}$  becomes
larger, and the closer to unity this transmission
amplitude becomes.

\section{Conclusions}  \label{CONCLUS}

We consider here two types modifications of the exactly-soluble
Demkov-Osherov model of a sloped linear potential curve crossing a set
of horizontal ones:

\noindent (a) Truncation of the domain of the model with the boundary
 conditions
specified at the truncation points.

\leftline{(b) Deformation of the sloped potential into a piecewise
-linear shape.}
\noindent These two modified problems are considered in the quasi
-degeneracy
approximation.  The main results of the present analysis are that the
transition amplitudes in both modified models are not to be
represented in the semiclassical form of a product of LZ amplitudes,
and that both models allow for counterintuitive transitions, which are
completely forbidden in semiclassical theories, as well as in the
original analytically-soluble Demkov-Osherov model.

\ack

This work was supported in part by grants from the US-Israel
Binational Science Foundation (PSJ and YBB) and by the U.S. Office of
 Naval
Research (PSJ).

%\begin{references}
     %JPB
%\end{references}
%\Figures
\begin{figure}

\psfig{figure=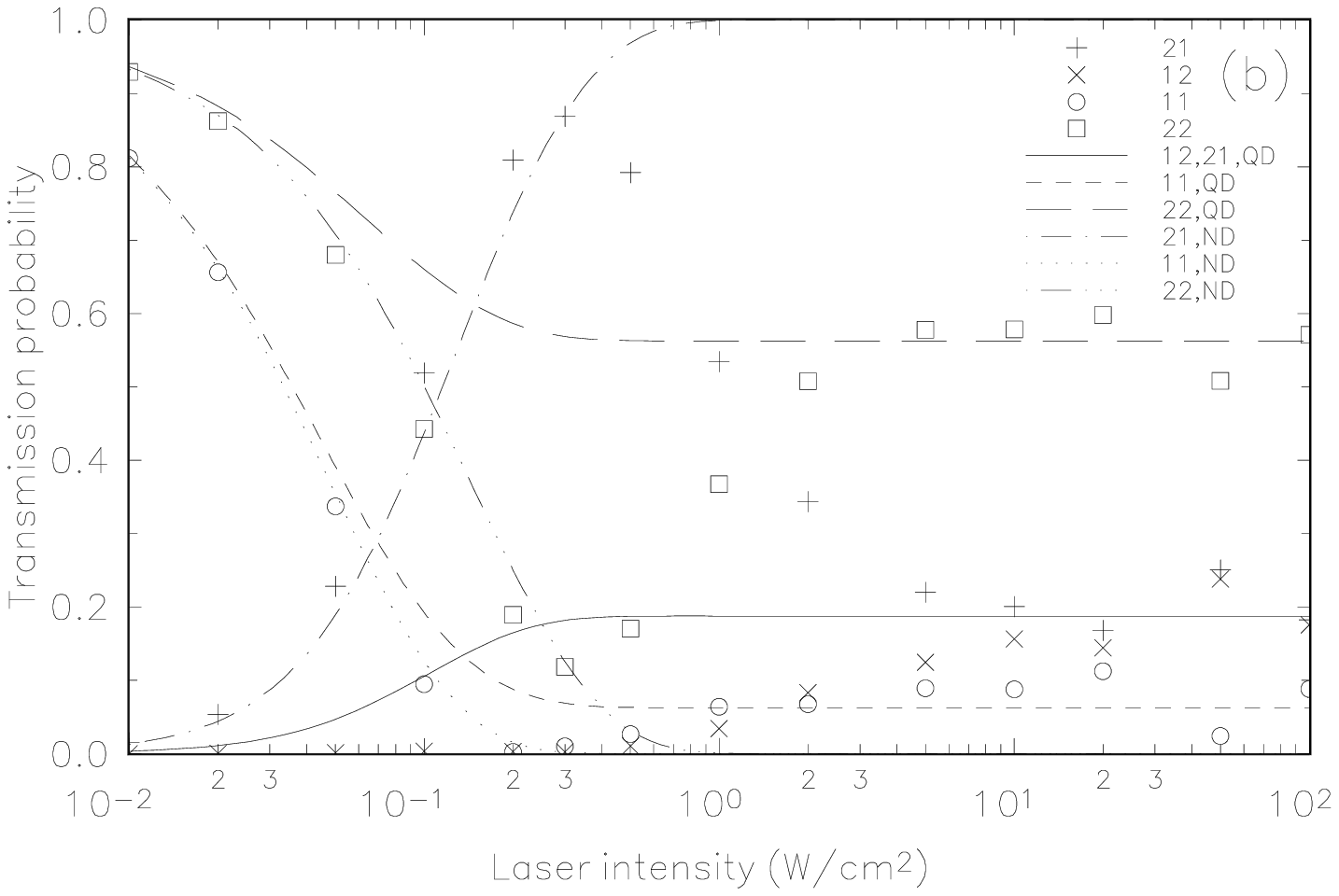}

\psfig{figure=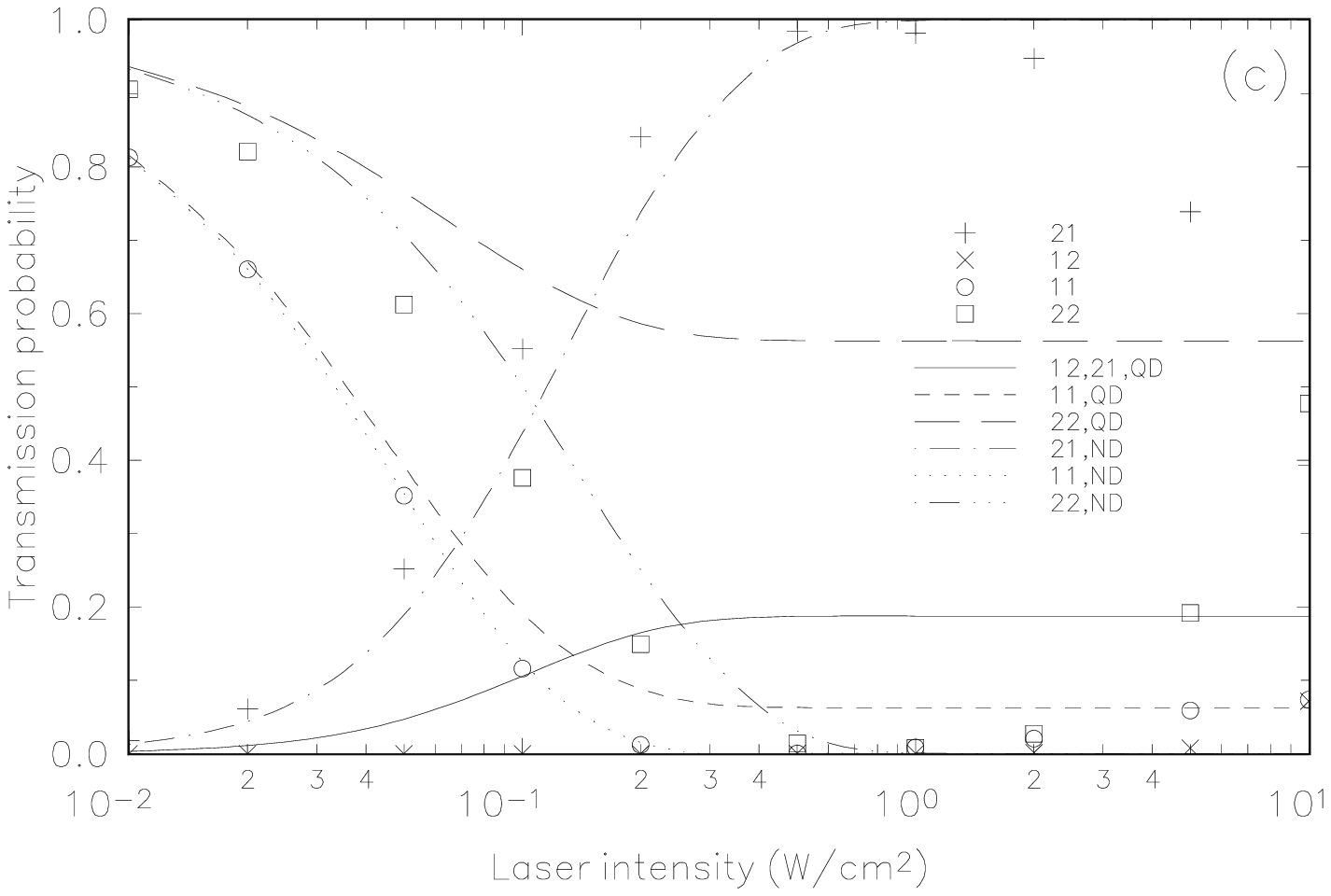}

\caption{Transmission probability $|S_{-1\nu ^\prime ,1\nu }|^{2}$
 for a truncated linear
problem of two horizontal potentials in one quasi-degenerate group.
 The
pairs of numbers assigned to the plots represent $\nu \nu ^\prime $.
 The coupling
constants are proportional to the square root of the laser intensity.
 The
curves are obtained using analytical expressions for the quasi
-degenerate
(QD) and non-degenerate (ND) cases. The points present results of the
numerical close-coupling calculations. The values of the gaps between
 the
horizontal potentials $V_{12}-V_{11}$  are ({\it a}) $6.7\times 10^{
-12}$, ({\it b})
$6.7\times 10^{-11}$, and ({\it c}) $5\times 10^{-10}$.}

\end{figure}
\begin{figure}

\psfig{figure=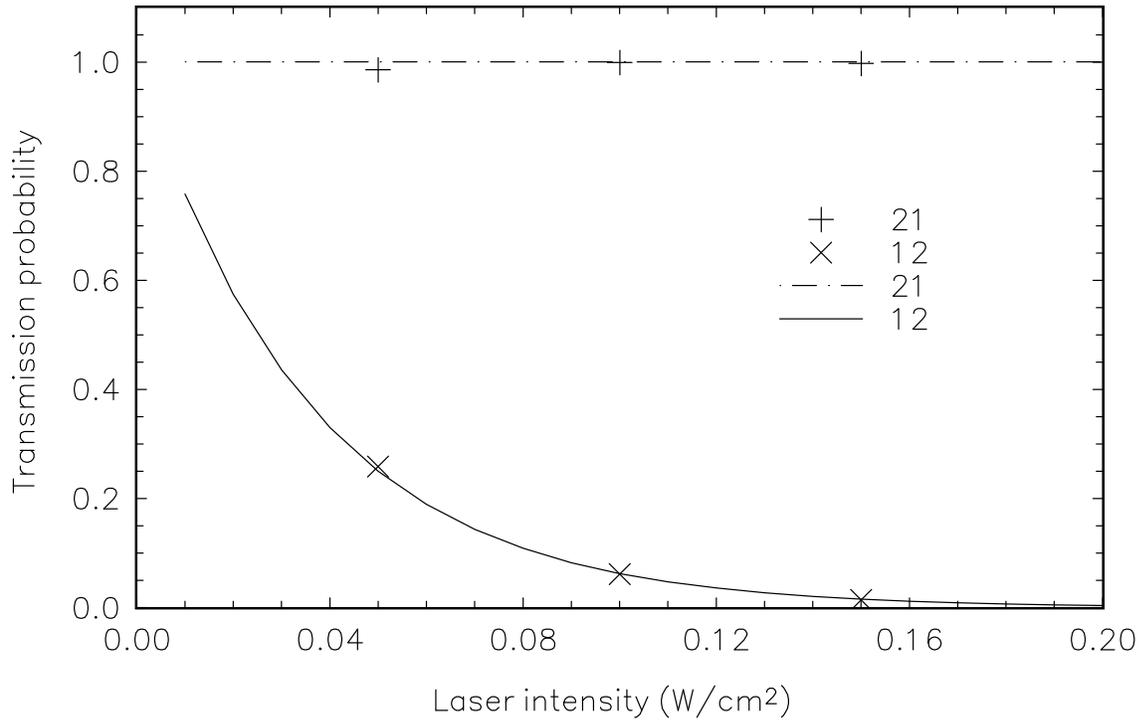}

\caption{Transmission probabilities $|S_{-1\nu ^\prime ,1\nu }|^{2}$
 for a piecewise-
linear model with horizontal wings, showing results of the analytical
expressions, in comparison with results of the numerical calculations
(represented by points). The pairs of numbers assigned to the plots
represent $\nu \nu ^\prime $.  The value of the gap between the
 horizontal potentials is
$V_{12}-V_{11}=6.7\times 10^{-13}$.}

\end{figure}
\end{document}